\documentclass[useAMS,usenatbib]{mn2e}
\usepackage{amsmath}
\usepackage{amssymb}
\usepackage{epsfig}
\usepackage{natbib}
\title[Evolution of the black hole scaling relations]{Towards an understanding of the evolution of the scaling relations for supermassive black holes.}
\author[C. M. Booth \& J. Schaye]{C. M. Booth$^{1}$\thanks{E-mail: booth@strw.leidenuniv.nl (CMB)} and Joop Schaye$^{1}$\\
$^{1}$Leiden Observatory, Leiden University, PO Box 9513, 2300 RA Leiden, the Netherlands}

\newcommand{\ion}[2]{\hbox{#1\,{\sc #2}}}

\newcommand{\civ}{\ion{C}{IV}}
\newcommand{\oiii}{\ion{O}{III}}

\newcommand{\epsf}{\epsilon_{\rm f}}
\newcommand{\epsr}{\epsilon_{\rm r}}
\newcommand{\mbh}{m_{\rm BH}}
\newcommand{\mhalo}{m_{\rm halo}}
\newcommand{\mseed}{m_{\rm seed}}
\newcommand{\reject}{r_{\rm ej}}
\newcommand{\rhalo}{r_{\rm halo}}

\newcommand{\ms}{m_\ast}
\newcommand{\us}{U_\ast}
\newcommand{\uhalo}{U_{\rm halo}}

\newcommand{\sigmas}{\sigma_\ast}
\newcommand{\msun}{{\rm M}_\odot}
\newcommand{\alphas}{\alpha_{\rm s}}

\newcommand{\asms}{0.52\pm0.05}
\newcommand{\asmh}{0.65\pm0.06}
\newcommand{\asss}{-0.32\pm0.05}
\newcommand{\asus}{0.05\pm0.06}
\newcommand{\asuh}{0.03\pm0.05}

\begin{document}
\pagerange{\pageref{firstpage}--\pageref{lastpage}} \pubyear{2010}
\maketitle
\label{firstpage}

\begin{abstract}
The growth of the supermassive black holes (BHs) that reside at the
centres of most galaxies is intertwined with the physical processes
that drive the formation of the galaxies themselves.  The evolution of
the relations between the mass of the BH, $\mbh$, and the properties
of its host therefore represent crucial aspects of the galaxy
formation process.  We use a cosmological simulation, as well as an
analytical model, to investigate how and why the scaling relations for
BHs evolve with cosmic time.  We find that a simulation that
reproduces the observed redshift zero relations between $\mbh$ and the
properties of its host galaxy, as well as the thermodynamic profiles
of the intragroup medium, also reproduces the observed evolution in
the ratio $\mbh/\ms$ for massive galaxies, although the evolution of
the $\mbh/\sigma$ relation is in apparent conflict with
observations. The simulation predicts that the relations between
$\mbh$ and the binding energies of both the galaxy and its dark matter
halo do not evolve, while the ratio $\mbh/\mhalo$ increases with
redshift. The simple, analytic model of \citet{boot10}, in which the
mass of the BH is controlled by the gravitational binding energy of
its host halo, quantitatively reproduces the latter two
results. Finally, we can explain the evolution in the relations
between $\mbh$ and the mass and binding energy of the stellar
component of its host galaxy for massive galaxies ($\ms\sim
10^{11}\,\msun$) at low redshift ($z<1$) if these galaxies grow
primarily through dry mergers.
\end{abstract}


\section{Introduction}
\label{sec:intro}

Over the past decade it has become clear that the supermassive black
holes (BHs) found at the centres of virtually all galaxies with
spheroidal components, have masses that are coupled to the properties
of their host galaxies
\citep{magg98,ferr00,trem02,hari04,hopk07b}. Additionally, there
exists evidence that BH masses are coupled to the properties of the
dark matter haloes in which they reside \citep{ferr02,boot10}.
Further correlations between quasar activity \citep[e.g.][]{boyl98}
and the evolution of the cosmic star formation rate
\citep[e.g.][]{mada96} provide evidence that there exists a link
between galactic star formation and accretion onto a central AGN.

It has long been recognised that the growth of BHs is likely
self-regulated \citep{silk98} and that these tight correlations
indicate that the growth of BHs is tightly intertwined with the
physical processes that drive galaxy formation.  However, despite a
wide variety of theoretical and observational studies, the origin of
these relations is still debated. The study of the evolution of the BH
scaling relations therefore represents a crucial aspect of the galaxy
formation process that may provide us with additional clues regarding
the physical processes that give rise to the BH scaling relations.

Addressing these questions observationally is challenging. Due to
their extremely high luminosities, bright quasars provide a promising
route to measuring BH masses at high redshift through the widths of
low-ionization lines that are associated with the broad-line region
close to the BH and using the assumption of virial equilibrium
\citep[e.g.][]{vest02}.  It has, however, been claimed that this
procedure systematically underestimates BH masses \citep{jarv02}.
Measuring galaxy masses for these objects is very difficult as the BH
outshines the galaxy by a large factor \citep[see e.g.\ the discussion  in ][]{merl09}. Since AGN surveys are biased towards
more massive black holes, selection effects also need to
be taken into account \citep[e.g.][]{shen09,benn10}, which can make it
difficult to distinguish between evolution in the normalization and
in the scatter in the scaling relations \citep{laue07}. In spite of these difficulties,
measurements of the BH scaling relations have been made as far out as
redshift three.

\citet{mclu06} found that the BHs associated with radio loud AGN
residing in galaxies of a given stellar mass are a factor of four more
massive at redshift two than in the local Universe.  \citet{deca09}
studied the \civ\, line associated with the quasar broad line region
in R-band selected hosts at both redshifts zero and three and found
that BHs are typically a factor of seven more massive at high redshift
for a given galaxy mass.  These results are consistent with other
observational studies
\citep{walt04,peng06,peng06a,merl09,gree10,benn10}. Taken together,
these papers suggest an emerging consensus that at higher redshift BHs
in hosts of a given mass are systematically more massive than in the
local Universe, although see \citet{jahn09} for one study that finds
no significant evolution.
 
The evolution of the relation between BH mass, $\mbh$, and stellar
velocity dispersion, $\sigmas$, has been studied utilising the width
of the \oiii\, line as a proxy for stellar velocity dispersion
\citep{nels96}.  These studies suggest that the $\mbh-\sigmas$
relation either does not evolve \citep{shie03,gask09}, or
does so weakly, with BHs $\sim 0.1-0.3$~dex more massive at $z=1$
\citep{salv06,gu09,woo08,treu07}.

The evolution of the BH scaling relations has also been studied using
numerical simulations \citep[e.g.][]{robe06,joha09} and semi-analytic
models \citep[e.g.][]{malb07,lama10,kisa10}. \citet{robe06} employed
simulations of idealised galaxy mergers, initialised to have
properties typical of merger progenitors at various redshifts, to
construct the relation between galaxy stellar mass, $\ms$, and
$\sigmas$ as a function of redshift and found that, at a given value
of $\sigmas$, the corresponding $\mbh$ decreases mildly with
increasing redshift.  At $z=1$ the simulations of \citet{dima08} have
BHs that lie slightly above the $z=0$ normalization of the
$\mbh-\sigma$ relation.  However, these simulations were stopped at
$z=1$ and so cannot inform us about the evolution of the
$\mbh-\sigmas$ toward lower redshift.  However, for $z>1$ they predict
a weak evolution in the $\mbh-\sigmas$ relation such that at higher
redshift galaxies of a given velocity dispersion contain slightly less
massive BHs.  \citet{joha09} employed similar numerical techniques to
argue that it is unlikely that BHs are able to form significantly
before their host bulges. Semi-analytic models that reproduce many
redshift zero properties of galaxies also predict that, at a fixed
$\sigmas$, BH masses decrease with increasing redshift
\citep{malb07}. These theoretical models thus predict evolutionary
trends that go in the opposite direction to those inferred from
observations.  Finally, the models of \citet{hopk09} predict that, at
a fixed stellar velocity dispersion, BH masses at higher redshift are
either the same (for $\mbh\sim10^8\,\msun$) or slightly more massive
(for $\mbh>10^8\,\msun$) at fixed $\sigmas$ than their redshift zero
counterparts, in agreement with observation.

On the other hand, the relation between BH mass and galaxy bulge mass
shows a positive evolution in both semi-analytic models \citep{malb07,
  hopk09} and numerical simulations \citep{dima08}, the magnitude of
which is comparable to that observed.  The larger spread in the
predictions for the evolution of the $\mbh-\sigmas$ relation may
reflect that it is more difficult to predict velocity dispersions,
which depend on both mass and size, than it is to predict masses.

In \citet[][hereafter BS09]{boot09} we presented self-consistent,
hydrodynamical simulations of the co-evolution of the BH and galaxy
populations that reproduce the redshift zero BH scaling
relations. These same simulations also match group temperature,
entropy and metallicity profiles, as well as the stellar masses and
age distributions of brightest group galaxies \citep{mcca10}.  In
\citet[]{boot10} (hereafter BS10) we used the same simulations, as
well as an analytic model, to demonstrate that $\mbh$ is determined by
the mass of the dark matter (DM) halo with a secondary dependence on
the halo concentration, of the form that would be expected if the halo
binding energy were the fundamental property that controls the mass of
the BH.  In the present work we use the same models to investigate why
and how the BH scaling relations evolve for massive galaxies.

This paper is organised as follows. In Sec.~\ref{sec:method} we
summarise the numerical methods employed in this study and the
simulation analysed.  In Sec.~\ref{sec:ev} we present predictions for
the evolution of the BH scaling relations and compare them to
observations. We find that the evolution in the $\mbh-\ms$ relation
predicted by the simulations is in excellent agreement with the
observations, while the measured weak evolution in the $\mbh-\sigma$
relation is in apparent disagreement, and predict that while BH mass
increases with redshift for fixed halo mass, the relations between
$\mbh$ and the binding energies of both the host galaxies and DM
haloes do not evolve. We demonstrate in \ref{sec:explanation} that the
a analytic description in which $\mbh$ is coupled to the DM halo
binding energy can reproduce the evolution of the relation between BH
and halo mass. Furthermore, we show that the evolution in the
relations between the BH and the stellar mass and binding energy can
be understood in terms of the more fundamental relation with the
binding energy of the dark halo and the growth of massive galaxies
through dry mergers.  Finally, we summarise our main conclusions in
Sec.~\ref{sec:conclusions}.

\section{Numerical Method}
\label{sec:method}
We have carried out a cosmological simulation using a significantly
extended version of the parallel PMTree-Smoothed Particle
Hydrodynamics (SPH) code {\sc gadget iii} \citep[last described in
][]{spri05b}.  The simulation and code are described in detail in
BS09, we provide only a brief summary here.  In addition to
hydrodynamic forces, we treat star formation \citep{scha08}, supernova
feedback \citep{dall08}, radiative cooling \citep{wier08},
chemodynamics \citep{wier09} and black hole accretion and feedback
\citep[BS09,][]{spri05}. We summarise in Sec.~\ref{sec:bhs} the
essential features of the BH model.

The properties of central galaxies and DM haloes are calculated by
first identifying the most gravitationally bound particle in each DM
halo using the algorithm {\sc subfind} \citep{spri01,dola09}, which is
then considered the halo centre.  All stars within a radius of
$0.15\rhalo$ are then assigned to the central galaxy.  We note that
our conclusions are insensitive to the exact choice for this radius,
and whether we use a fixed physical value or a fixed fraction of the
halo virial radius. As long as the sphere encloses the central object,
our results are insensitive to this choice.  Halo mass, $\mhalo$, is
calculated as the total mass enclosed within a sphere, centred on the
most bound particle in the halo, that has a mean density of 200 times
the mean density of the Universe and the virial radius $\rhalo$ is the
radius of this sphere.  Because it is not expected that the same
physics holds for both the central galaxy of a halo and its
satellites, which are expected to rapidly have their gas supply
stripped when they become satellites, our analysis is restricted to
BHs identified as residing in the central galaxy in a DM halo, defined
as the galaxy closest to the centre of the DM potential well of each
halo.

\subsection{The black hole model}
\label{sec:bhs}
Seed BHs of mass $m_{{\rm seed}}=10^{-3}m_{{\rm g}}\approx
10^5\,\msun$ -- where $m_{{\rm g}}$ is the simulation gas particle
mass -- are placed into every DM halo that contains more than 100 DM
particles (which corresponds to a DM halo mass of
$4.1\times10^{10}\,\msun/h$) and does not already contain a BH
particle.  Haloes are identified by regularly running a
friends-of-friends group finder during the simulation.  After forming,
BHs grow by two processes: accretion of ambient gas and mergers.  Gas
accretion occurs at the minimum of the Eddington rate, $\dot{m}_{{\rm
    Edd}}=4\pi Gm_{{\rm BH}} m_{\rm p}/\epsr \sigma_{{\rm T}} c$ and
$\dot{m}_{{\rm accr}}=\alpha4\pi G^2 m_{{\rm BH}}^2
\rho/(c_{s}^2+v^2)^{3/2}$, where $m_{\rm p}$ is the proton mass,
$\sigma_{\rm T}$ is the Thomson cross-section, $\epsr$ is the
radiative efficiency of the BH, $c$ is the speed of light, $c_{s}$ and
$\rho$ are the sound speed and gas density of the local medium, $v$ is the
velocity of the BH relative to the ambient medium, and $\alpha$ is a
dimensionless efficiency parameter.  The parameter $\alpha$ accounts
for the fact that our simulations possess neither the necessary
resolution nor the physics to accurately model accretion onto a BH on
small scales.  Note that for $\alpha=1$ this accretion rate reduces to
the so called Bondi-Hoyle \citep{bond44} rate.

As long as we resolve the scales and physics relevant to Bondi-Hoyle
accretion, we should set $\alpha=1$.  If a simulation resolves the
Jeans scales in the accreting gas, then it will also resolve the
scales relevant for Bondi-Hoyle accretion onto any BH larger than the
simulation mass resolution (BS09).  We therefore generally set
$\alpha$ equal to unity.  However, this argument breaks down in the
presence of a multi-phase interstellar medium, because our simulations
do not resolve the properties of the cold, molecular phase, and as
such the accretion rate may be orders of magnitude higher than the
Bondi-Hoyle rate predicted by our simulations for star-forming gas.
We therefore use a power-law scaling of the accretion efficiency such
that $\alpha=(n_{{\rm H}}/n_{{\rm H}}^*)^\beta$ in star-forming gas,
where $n_{{\rm H}}^*=0.1\,{\rm cm}^{-3}$ is the critical density for
the formation of a cold, star-forming gas phase.  The parameter
$\beta$ is a free parameter in our simulations.  We set $\beta=2$, but
note that the results shown here are insensitive to changes in this
parameter when $\beta \ga 2$.  We note that we do not resolve the
Bondi radius of BHs less massive than the particle mass in our
simulations, and that for BHs with $\mbh\sim m_{\rm g}$ the Bondi
radius is unresolved unless the density is low or the temperature
high.  Our choice of $\alpha=1$ therefore provides an underestimate of
the true accretion rate in these regimes.  However, even setting
$\alpha=100$ for all densities gives very similar results
\citep{boot09}.  This is because all BHs accrete almost all of their
mass in short, (near) Eddington-limited bursts of accretion and thus
that our treatment of accretion in low-density environments is less
important.  The accretion model in high-density environments is
necessarily very crude, but we note that our results are insensitive
to the details of the accretion model as long as $\alpha$ is
sufficiently large that the BHs become more massive than observed in
the absence of feedback and if two reasonable conditions are met that
are necessary for self-regulation to be possible.  Firstly, the BH
accretion rate must increase with increasing density, and secondly it
must increase with BH mass (see BS09).

Energy feedback is implemented by allowing BHs to inject a fixed
fraction of the rest mass energy of the gas they accrete into the
surrounding medium.  The energy deposition rate is given by
\begin{equation}
\dot{E}=\epsf\epsr\dot{m}_{\rm accr}c^2=\frac{\epsf\epsr}{1-\epsr}\dot{m}_{\rm
BH}c^2\,,
\label{eq:edot}
\end{equation}
where $\dot{m}_{\rm accr}$ is the rate at which the BH is accreting
gas, and $\dot{m}_{\rm BH}$ is the rate of BH mass growth.

We set $\epsr$ to be 0.1, the mean value for radiatively efficient
accretion onto a Schwarzschild BH \citep{shak73} and use $\epsf=0.15$
as our fiducial value.  It was shown in BS09 that, for $\epsf=0.15$,
this simulation reproduces the observed redshift zero $\mbh-\ms$ and
$\mbh-\sigmas$ relations.  Energy is returned to the surroundings of
the BH \lq thermally\rq, by increasing the temperature of $N_{\rm
  heat}$ of the BH's neighbouring SPH particles by at least $\Delta
T_{\rm min}$.  A BH performs no heating until it has built up enough
of an energy reservoir to heat by this amount. Imposing a minimum
temperature increase ensures that the radiative cooling time is
sufficiently long for the feedback to be effective.  We set $N_{\rm
  heat}=1$ and $\Delta T_{\rm min}=10^8$\,K but the results are
insensitive to the exact values of these parameters (see BS09).

\subsection{The cosmological simulation}
The simulation employed in the current work uses a cubic box of size
50 comoving Mpc/$h$ and assumes periodic boundary conditions. The
simulation contains $256^3$ particles of both gas and collisionless
cold DM and is evolved down to redshift zero.  The DM and initial
baryonic particle masses are $4.1\times10^8\,\msun/h$ and
$8.7\times10^7\,\msun/h$, respectively.  Comoving gravitational
softenings are set to $1/25$ of the mean interparticle separation down
to $z=2.91$, below which they switch to a fixed proper scale of
$2\,{\rm kpc}/h$.  The simulation employed in this study was
previously also analysed as the fiducial simulation in BS09 and BS10

Comparison of the simulation employed in this study to an otherwise
identical one with eight (two) times lower mass (spatial) resolution
informs us in what mass and redshift range the relations between
$\mbh$ and galaxy and halo properties are numerically converged.  The
relation between $\mbh$ and both $\ms$ and $\mhalo$ is numerically
converged up to redshift two for all haloes with $\mbh>10\,\mseed$.
Measurements of stellar velocity dispersion are, however, only
converged for $z<1$ and $\mbh > 10^2\,\mseed$.  We will only give
results for redshifts and BH masses for which the results are
converged with respect to numerical resolution.

We note that these simulations do not resolve the scales on which the
BH is the gravitationally dominant component in the galaxy and so
cannot be used to study BH self-regulation on the smallest scales.
However, the simulations do have sufficient resolution for baryons to
be gravitationally dominant in the centres of haloes.  We cannot
conclusively rule out that if we increased our mass resolution
significantly and used more sophisticated sub-grid models that the BH
would self-regulate on different scales.  However, suggestively, in
\citet{boot10} we verified that in simulations with a spatial
resolution of 0.5~kpc/$h$ that, at $z=2$, the BH masses scale in the
same way as in the lower-resolution simulations.

\begin{figure}
\begin{center}
\includegraphics[width=8.3cm,clip]{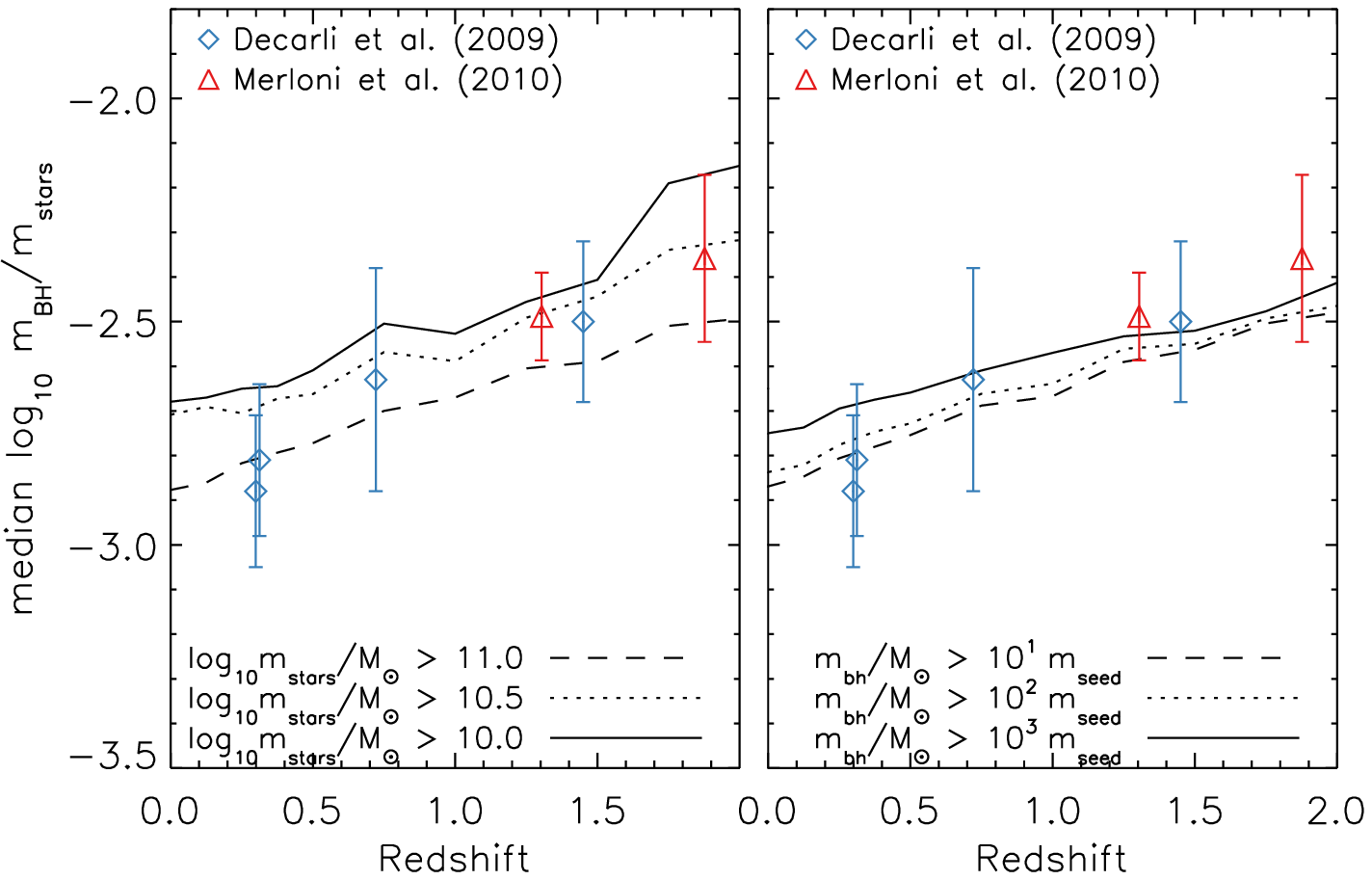}
\end{center}
\caption{The median $\mbh/\ms$ ratio as a function of redshift for
  galaxies of different stellar masses (\emph{left panel}) and for BHs
  above a given mass (\emph{right panel}) predicted by the
  cosmological simulation.  The points with error bars show the
  measurements of \citet[][red triangles]{merl09} for a sample of
  galaxies in the mass range $10.5<\log(\ms/\msun)<11.5$ and
  \citet[][blue diamonds]{deca09} for a sample of galaxies with
  typical stellar masses $\sim10^{11}\,\msun$.  We show a number of
  different mass cuts for the simulation results to demonstrate that
  the results are insensitive to the particular mass cut chosen.
  Both the simulations and observations in this plot show the
  \emph{total} stellar mass of the galaxies. The simulation predicts   that the median $\mbh/\ms$ ratio
  increases with redshift. The predictions are in excellent agreement
  with the observations, regardless of the particular mass cut made.}
\label{fig:ms}
\end{figure}

\section{Simulation results}
\label{sec:ev}
\subsection{The evolving relations between black holes and galaxies}
\label{sec:bhgal}
Fig.~\ref{fig:ms} compares the predicted evolution of the median
$\mbh/\ms$ ratio for different minimum stellar (left panel) and BH
(right panel) masses with observations of AGN in galaxies with
$\ms\sim 10^{11}\,\msun$. The $\mbh/\ms$ ratio increases with
redshift, in close agreement with observations \citep{merl09,deca09}.
At redshift zero this agreement is unsurprising because the efficiency
of AGN feedback in the simulation was tuned to reproduce the
normalisation of the $z=0$ BH scaling relations.  The agreement at
higher redshift represents, however, a non-trivial prediction of a
model in which BHs self-regulate their accretion through the coupling
of a small fraction of the radiative energy to the ambient medium.

\begin{table}
\caption{Evolution of the normalisation of the power-law relations
  between BH mass and the properties of its host. The sample includes
  all galaxies that contain a BH with
  $\mbh>10^2\,\mseed\approx10^7\,\msun$, corresponding to 162 (132)
  haloes at $z=0$ ($z=1$). The median BH, stellar, and halo mass are $10^{7.6}$, $10^{10.6}$, and $10^{12.8}\,\msun$, respectively. The central column gives the median change
  in $\log_{10}\mbh$ at redshift one relative to $z=0$ for fixed
  values of the quantity listed in the left column. The right column
  shows the slope, $\alphas$, of the best fit power-law describing the
  rate of evolution of the scaling relation (Eq.~\ref{eq:alpha}) over
  the redshift range $0-1$.  From top to bottom, we consider evolution
  of $\mbh$ for fixed stellar mass, halo mass, central stellar velocity dispersion, galaxy binding energy
  ($\ms\sigmas^2$), and halo binding energy (Eq.~\ref{eq:be}).  Errors
  are calculated from $10^3$ bootstrap resamplings of the data.}
\begin{center}
\begin{tabular}{r|r|r}
Variable & $\Delta\log_{10}\frac{\mbh(z=1)}{\mbh(z=0)}$ & $\alphas$\\
\hline
$\ms$             & $0.20\pm0.05$ & $\asms$\\
$\mhalo$          & $0.23\pm0.03$ & $\asmh$\\
$\sigmas$         &$-0.09\pm0.04$ & $\asss$\\
$U_\ast$          & $0.02\pm0.03$ &  $\asus$\\
$U_{\rm halo}$     & $0.01\pm0.03$ &  $\asuh$\\
\end{tabular}
\end{center}
\label{tab:deltas}
\end{table}

In Table \ref{tab:deltas} we show the predicted evolution up to $z=1$
of the amplitude of the relations between $\mbh$ and the masses,
velocity dispersions and binding energies of both the host galaxies
and the host DM haloes.  The central column gives
$\log_{10}\mbh(z=1)-\log_{10}\mbh(z=0)$, calculated by fitting
power-law relations under the assumption that the slopes of the
scaling relations do not evolve, which is a good approximation in the
redshift range studied here\footnote{At $z=0$ ($z=1$) the slope of the
  relation between $\mbh$ and $\uhalo$ is $1.01\pm0.14$
  ($0.96\pm0.17$) and the slope of the relation between $\mbh$ and
  $\us$ is $0.93\pm0.07$ ($0.96\pm0.09$), consistent with the $z=0$
  observational results of \citet{feol07} and \citet{feol10}. At all
  redshifts these slopes are consistent with unity.  The slopes of the
  relations between $\mbh$ and $\ms$, $\sigmas$ and $\mhalo$ are
  $1.16\pm 0.06$ ($1.2\pm 0.2$), $4.6\pm0.8$ ($4.4\pm 0.8$) and
  $1.5\pm0.2$ ($1.5\pm0.3$). There is thus no evidence for evolution
  in any of the slopes, which agrees with the results of
  \citet{robe06} for the $\mbh - \sigmas$ relation and with
  \citet{dima08} for the $\mbh - \ms$ relation. The slopes of the
  $z=0$ relations are consistent with observations
  \citep{hari04,trem02,band09}.}.  The right-most column of Table
\ref{tab:deltas} gives the slope, $\alphas$, of the power-law
evolution in the amplitude of each scaling relation
\begin{equation}
\label{eq:alpha}
\frac{\mbh}{X^{n_0}}\propto(1+z)^{\alphas}\,,
\end{equation}
where $X$ is one of the variables listed in the left column, and
$n_{0}$ is the slope of the $\mbh-X$ relation at $z=0$.  We find that
$\alphas=\asms$ for the $\mbh-\ms$ relation, in good agreement
with \citet{dima08}, who found $\alphas=0.5$.

The evolution of the $\mbh-\sigmas$ relation is smaller but
significant, with $\alphas=\asss$. This is in apparent disagreement
with various observational studies that either infer a positive
\citep{salv06,treu07,woo08,gu09} or negligible \citep{shie03,gask09}
evolution in the normalisation of the $\mbh-\sigmas$ relation.  The
predicted evolution does, however, agree with other simulation studies
\citep{robe06} and semi-analytic models \citep{malb07}. Taken
together, the simulation predicts that the $\mbh-\ms$ and
$\mbh-\sigmas$ relations evolve such that the relation between $\mbh$
and stellar binding energy ($\propto\ms\sigmas^2$) is independent of
redshift ($\alphas=\asus$).

It is tempting to conclude from the finding that the ratio
$\mbh/(\ms\sigmas^2)$ does not evolve that the BH mass is determined by
the binding energy of the galaxy. However, we demonstrated explicitly
in BS10 that the BH mass is instead controlled by the binding energy
of the DM halo. This implies that the binding energy of the galaxy
tracks the binding energy of the halo, which we will confirm and
explain below.

\subsection{The evolving relations between black holes and dark matter haloes}
\label{sec:bh-halo}
We now turn our attention to the relations between the mass of the BH
and the DM halo in which it resides.  In BS10 we argued that a BH
grows until it has injected an amount of energy into its surroundings
that scales with the binding energy of its host DM halo.  We therefore
do not expect the $\mbh-U_{\rm halo}$ relation to evolve. Indeed, in
the simulation the amplitude of this relation is independent of
redshift, with $\alphas=\asuh$, as would be expected for a fundamental
link between $\mbh$ and the binding energy of the host DM halo.

\begin{figure}
\begin{center}
\includegraphics[width=8.3cm,clip]{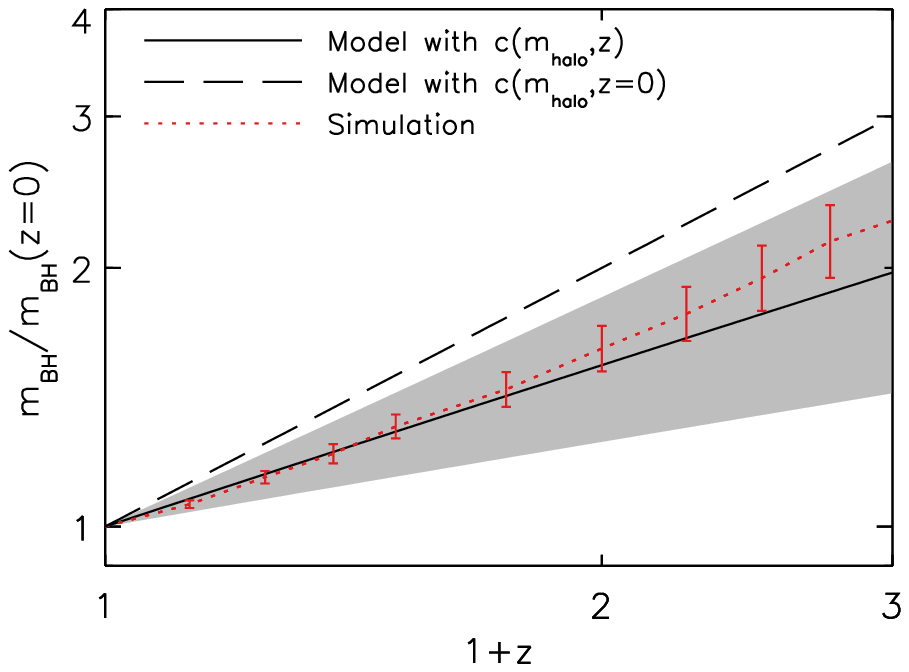}
\end{center}
\caption{Predicted evolution of the normalisation of the $\mbh-\mhalo$
  relation.  The red, dotted curve shows the median evolution
  predicted by the cosmological simulation when all BHs for which both
  BH and halo properties are well resolved ($\mbh>10^2\,\mseed$) are
  included. The grey region represents the allowed range in evolution
  predicted by the analytic model of BS10, which assumes that the BH
  mass is controlled by the binding energy of the DM halo. The binding
  energy of an NFW halo depends on mass, redshift, concentration
  (which itself depends on both mass and redshift) and on the radius,
  $\reject$, at which it is evaluated.  The grey region corresponds to
  $\reject/\rhalo=0.1-1.0$ (bottom to top) and the solid black line to
  $\reject=0.22\,\rhalo$, the value for which we which we predict
  $\mbh\propto\mhalo^{1.55}$ at $z=0$, in accord with both
observations
  (Bandara et al. 2009) and simulation (BS10). For comparison, the
  black dashed line shows the evolution that our analytic model would
  have predicted if we had ignored the evolution of the $c(\mhalo$)
  relation.  At all redshifts the normalisation of the simulated
  $\mbh-\mhalo$ relation (red, dotted curve) agrees with that
  predicted by the analytic model based on the assumption that the
  fundamental relation is between BH mass and the binding energy of
  the DM halo.}
\label{fig:mh}
\end{figure}

We do, however, expect the relation between $\mbh$ and $\mhalo$ to
evolve. At higher redshift, haloes of a given mass are more compact
than their redshift zero counterparts and are thus more strongly
gravitationally bound.  This means that, at a fixed halo mass, more
energy is required to eject gas from haloes at high redshift and in
order to self-regulate, BHs must grow to be more
massive. Fig.~\ref{fig:mh} shows the normalisation of the
$\mbh-\mhalo$ relation as a function of redshift and confirms that the
simulation predicts the amplitude of this relation to increase with
redshift (red, dashed curve), with $\alphas = \asmh$.

\section{Explaining the evolution}
\label{sec:explanation}

As we already noted, the idea that the binding energy of the dark halo controls the mass of the BH explains our finding that the $\mbh-U_{\rm halo}$ relation is independent of redshift. We will now show that the analytic model of BS10 also reproduces the evolution of the $\mbh-\mhalo$ relation and that it can explain the observed evolution in the scaling relations with the stellar properties if the observed galaxies evolve predominantly through dry mergers, as predicted by the simulation.

\subsection{The $\mbh-\mhalo$ relation}

If the
energy injected by a BH is proportional to the halo gravitational
binding energy, $U_{\rm halo}$, then, for a DM halo with an NFW
\citep{nava97} density profile (BS10)
\begin{align}
\mbh\propto U_{\rm halo} \propto \frac{\mhalo^{2}}{\rhalo}\propto f(c,x)(1+z)\mhalo^{5/3}\,,
\label{eq:be}
\end{align}
where $\mhalo$ is the halo mass, $c(\mhalo,z)$ is the halo
concentration, $x$ is defined to be $x\equiv\reject/\rhalo$, $\reject$
is the physical scale on which BH self-regulation takes place, and
$f(c,x)$ is the function
\begin{align}
f(c,x)=&\frac{c}{\big(\ln(1+c)-c/(1+c)\big)^2} ~ \times \nonumber\\
&\Bigg(1-\frac{1}{(1+cx)^2}-\frac{2\ln(1+cx)}{1+cx}\Bigg)\,.
\label{eq:f}
\end{align}
Simulations have shown that $c$ is a function of both redshift and
halo mass, and scales approximately as\footnote{Note that the slope of
  the power-law dependence of concentration on redshift depends on the
  halo definition used \citep{duff08}.} \citep{duff08}
\begin{equation}
\label{eq:c}
c\propto\mhalo^{-0.1}(1+z)^{-0.5}\,.
\end{equation}
Combining Eqs.~\ref{eq:be}\,-\,\ref{eq:c}, BS10 found the slope of the
$\mbh-\mhalo$ relation to be weakly dependent on $\reject$. At $z=0$
it varies from $n_0=1.50$ for $x=0.1$ to $n_0=1.61$ for $x=1.0$.  In
order to exactly match the slope of 1.55 that is both observed
(\citealt{band09} find $1.55 \pm 0.31$) and predicted by the
simulations (BS10 find $1.55 \pm 0.05$ for the same simulation as is
analyzed here\footnote{We quoted a slope of $1.5 \pm 0.2$. Our error
  bar is greater because BS10 used $\mbh > 10 \mseed$ whereas we
  require $\mbh > 100 \mseed$.}), we would need to use $x=0.22$.

By using an NFW density profile and Eq.~(\ref{eq:c}), we have
implicitly assumed that the dark matter profile is well described by
the results obtained from simulations that include only dark
matter. \citet{duff10} have recently shown that, on the scales of
interest here, the back-reaction of the baryons onto the dark matter
is in fact very small if feedback from AGN is included, as required to
reproduce the observed stellar and gas properties of groups of
galaxies \citep{mcca10,puch08,fabj10,duff10}.

If, as argued in BS10, the BH mass is controlled by the DM halo
binding energy, then we expect the $\mbh-\mhalo$ relation to
evolve because the halo binding energy depends not only on halo
mass, but also on the virial radius and concentration, both of which
vary with redshift for a fixed halo mass.

If the $c-\mhalo$ relation did not evolve, then we would expect
$\mbh(\mhalo)\propto (1+z)$ (Eq.~\ref{eq:be}).  However,
because halo concentration decreases with redshift (Eq.~\ref{eq:c}) we
expect the actual evolution of the $\mbh-\mhalo$ relation to be
weaker, i.e.\ $\alphas<1$.  The resulting relation between BH mass and
DM halo binding energy predicts that, at a given $\mhalo$, $\mbh$
increases with redshift and that by $z=2$ BHs are between 1.5 (for
$\reject/\rhalo=0.1$) and 2.6 (for $\reject/\rhalo=1.0$) times more
massive than at redshift zero.  For our
fiducial radius of self-regulation of $x=0.22$, BHs are 2.1 times
more massive, in excellent agreement with the simulation prediction of
$\alphas=\asmh$ (Table~\ref{tab:deltas}).

The evolution predicted by Eqs.~\ref{eq:be}\,-\,\ref{eq:c} is shown in
Fig.~\ref{fig:mh}.  The grey shaded region outlines the analytic
prediction for the evolution in BH mass over the range
$\reject/\rhalo=0.1-1.0$ and the solid black line shows the prediction
for $\reject/\rhalo=0.22$ (the value that reproduces the slope of the
redshift zero $\mbh-\mhalo$ relation). The red, dotted curve shows the
simulation prediction for the evolution of the $\mbh-\mhalo$ relation,
including all BHs with $\mbh>10^2\,\mseed$.  At all redshifts the
normalisation of the simulated $\mbh-\mhalo$ relation is compatible
with that predicted by the analytic model. For comparison, the dashed,
black line shows the predicted evolution of the $\mbh-\mhalo$ relation
if $c(\mhalo)$ were independent of redshift. The analytic model can
only reproduce the simulation result if the evolution of the
concentration-mass relation is taken into account.

The evolution of the $\mbh-\mhalo$ relation thus provides additional
evidence for the idea that the masses of BHs are determined by the
binding energies of the haloes in which they reside. 

\subsection{The relations between $\mbh$ and galaxy stellar properties}

\begin{figure}
\begin{center}
\includegraphics[width=8.3cm,clip]{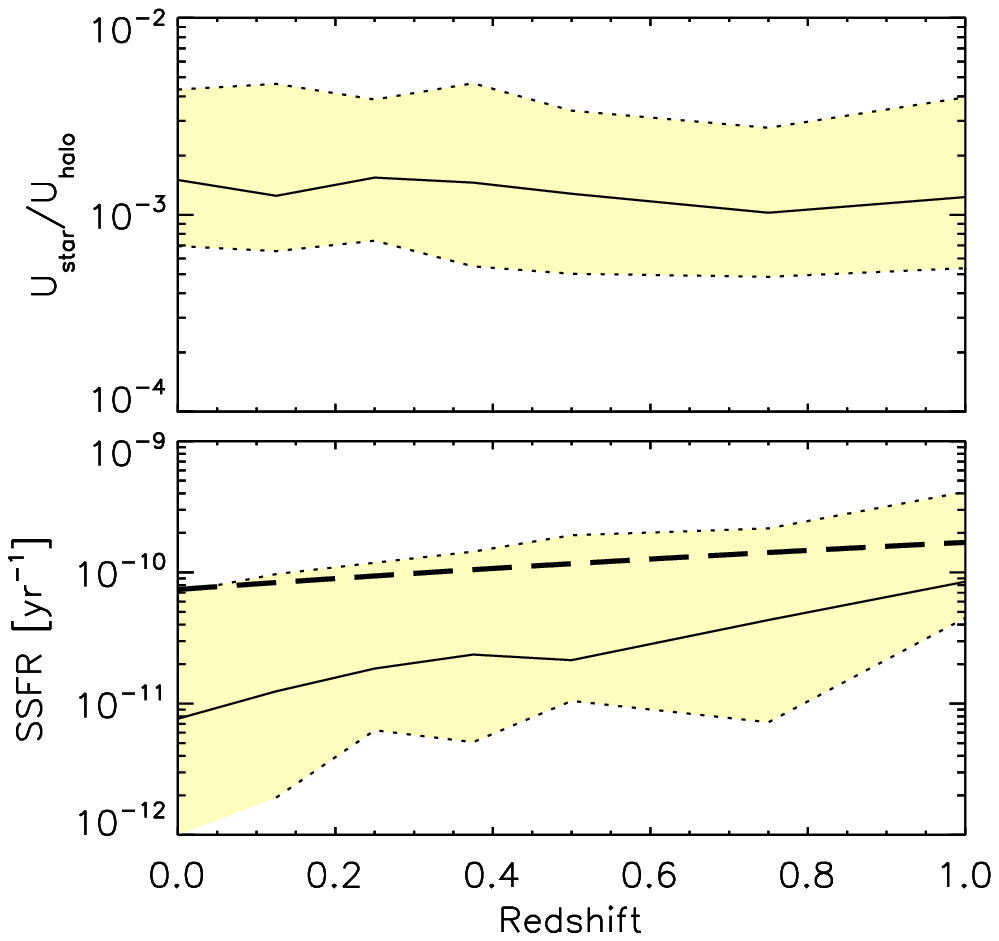}
\end{center}
\caption{Galaxy properties as a function of redshift for $\ms \approx
  10^{11}\,\msun$ (at each redshift we selected the 20 galaxies with
  stellar masses closest to this value).  The yellow shaded region
  shows the area that contains the 25th to 75th percentiles of the
  data and the black, solid curve shows the median. The top panel
  shows the ratio of stellar to halo binding energy and the bottom
  panel shows the specific star formation rate
  (SSFR$\equiv\dot{m}_\ast/\ms$). The dashed line in the bottom panel
  shows the inverse of the age of the Universe. Galaxies with SSFRs
  below this line may be considered passive.  }
\label{fig:tracks}
\end{figure}

Considering now only the stellar masses for which the evolution has
been measured observationally ($\ms\sim10^{11}\,\msun$) and the
redshift range for which all of the stellar and BH properties of the
galaxies are converged numerically ($z<1$), we ask if we can explain
how the relations between BH mass and galaxy stellar properties
evolve.  BS10 showed that the BH mass is determined by the binding
energy of the DM halo, which explains why the $\mbh-\uhalo$ relation
does not evolve. We find that the $\mbh-\us$ relation also does not
evolve, implying that, over the range of redshifts and masses
investigated here, $\us\propto\uhalo$. The top panel of
Fig.~\ref{fig:tracks} confirms that this is indeed the case in our
simulation.

For the binding energy of the galaxy to track that of the halo, we
require the two to grow through the same mechanism. This condition is
met if the galaxies grow primarily through dry mergers. In the absence
of significant in-situ star formation, both the stellar component,
which is predominantly spheroidal for massive galaxies, and the DM
halo are collisionless systems and are therefore expected to evolve in
a similar manner.

The bottom panel of Fig.~\ref{fig:tracks} shows that at $z\ll 1$ the
specific star formation rates (SSFR$\equiv\dot{m}_{\rm s}/\ms$) of
galaxies with $\ms \sim 10^{11}\,\msun$ are significantly lower than
the inverse of the Hubble time, $1/t_{\rm H}$, implying that the
galaxies are indeed not growing significantly via in-situ
star-formation, in agreement with various observations
\citep[e.g.][]{schaw06,vdwel09,beza09,vdok10}.  The analysis in
Fig.~\ref{fig:tracks} was carried out for galaxies at a fixed stellar
mass, but the same results hold if we trace individual galaxies
through time.  While the stellar masses of the most massive $z=0$
galaxies grew on average by a factor 2.65 since $z=1$, the fraction of
stars in the redshift zero objects with birth redshifts below $z=1$
is, on average, only $15$\%.

\citet{merl09} studied the evolution of the BH scaling relation in the
redshift range $1<z<2.2$, and found that at $z=1$ a large fraction of
galaxies would be identified as star-forming. This is consistent with
our results as the SSFRs of our galaxy sample are increasing with
increasing redshift so that a significant fraction of them would be
identified as star-forming at $z\approx 1$. The fraction would be even
higher if we had required the galaxies to contain active AGN, as is
the case for the objects selected in the observations, because AGN
activity is typically accompanied by a temporary increase in the star
formation rate.

One might naively think that the $\mbh-\ms$ relation should not evolve
at all if the galaxies grew predominantly through dry
mergers. However, the progenitors of galaxies with $\ms\sim
10^{11}\,\msun$ at $z=0$ typically formed their stars and grew their
BHs later than the progenitors of galaxies that already have the same
stellar mass at $z=1$. Because the progenitors of higher redshifts
galaxies formed earlier, they have higher binding energies and thus
greater BH masses relative to their halo masses. Thus, even if $\ms
\sim 10^{11}\,\msun$ galaxies are growing predominantly by dry mergers
at both redshifts, they may have different BH masses. Observe,
however, that the evolution in the $\mbh-\ms$ relation that we predict
for these massive galaxies is only mild (about 0.2~dex; see
Fig.~\ref{fig:ms}) and that in situ star formation thus become
important for $z\ga 1$ (see Fig.~\ref{fig:tracks}).

If the ratio of the stellar to halo binding energies remains constant,
we can write
\begin{equation}
\mbh\propto\uhalo\propto\us\propto\ms\sigmas^2\,.
\label{eq:props}
\end{equation}
To explain the evolution of $\mbh/\ms$ that is observed for $\ms\sim
10^{11}\,\msun$, and which the simulation reproduces, we need to know
how the $\ms-\sigmas$ relation evolves for such galaxies.

Measurements of the evolution of the $\ms-\sigmas$ relation have so
far only been undertaken for a small number of objects. \citet{capp09}
presented stacked observations of seven galaxies with
$\ms\sim10^{11}\,\msun$ in the redshift range $1.6< z< 2.0$ and found
that these galaxies typically have the same $\sigmas$ as the very
highest velocity dispersion early-type galaxies of the same mass in
the local Universe. This is in agreement with observations showing
that galaxies of a given stellar mass are more compact at higher
redshifts \citep[see e.g.][]{will09}. Indeed, for early-type galaxies
with stellar masses $\sim 10^{11}\,\msun$ \citet{cena09} find that
typical velocity dispersions decrease from $\approx 240\,$~km/s at
$z=1.6$ to $\approx 180$~km/s at $z=0$, which implies $\sigmas\propto
(1+z)^{0.3}$. Combined with Eq.~(\ref{eq:props}) this yields $\mbh/\ms
\propto (1+z)^{0.6}$, in good agreement with the $\alphas\approx 0.5$
that is observed for galaxies of this mass.

\section{Conclusions}
\label{sec:conclusions}

We have used a self-consistent cosmological simulation
that reproduces the observed redshift zero relations between $\mbh$
and both galaxy and halo properties (as well as the thermodynamic
profiles of the intragroup medium) to investigate how, and why, these
relations evolve through time.  

The relation between BH mass and host galaxy mass predicted by the
simulation is consistent with available observations at $z<2$, which
are currently confined to $\ms\sim 10^{11}\,\msun$ galaxies. For such
galaxies we predict that the ratio $\mbh/\ms \propto (1+z)^{\alphas}$,
with $\alphas \approx 0.5$, and $\mbh/\sigmas^4 \propto (1+z)^{\alphas}$
with $\alphas\approx-0.3$, in apparent conflict with recent
observations.  The ratio between the BH mass and the binding energy of
the dark halo is independent of redshift, in agreement with BS10 who
argued that the BH mass is controlled by the halo binding energy.  The
simple analytic model of BS10, in which the BH mass is assumed to
scale in proportion to the binding energy of the dark halo, not only
reproduces the simulated redshift zero $\mbh-\mhalo$ relation, but
also its evolution. For a fixed halo mass BHs are more massive at
higher redshift because the haloes are more compact and thus more
tightly bound. Assuming an NFW halo density profile and the evolution
of the halo concentration-mass relation predicted by simulations, the
model can quantitatively account for the predicted evolution.

The simulation predicts that the ratio between the BH mass and the
binding energy of the stellar component of the galaxy is also
independent of redshift (at least for $\ms\sim10^{11}\,\msun$ and
$z<1$), even though BS10 demonstrated explicitly that the correlations
between BH mass and stellar properties are not fundamental. This
result is, however, consistent with a picture in which massive
galaxies grow primarily through dry mergers at low redshift, which we
showed to be the case in the simulation. Combined with the observed
evolution in the $\ms-\sigmas$ relation, this idea can quantitatively
account for the evolution in the $\mbh-\ms$ relation.

One interesting implication of this scenario is that the evolution of
the relations between BHs and the properties of their host galaxies
may differ for galaxies that do not grow predominantly through dry
mergers, as would be expected for lower masses and at higher
redshifts. We will investigate this further in the a future work,
employing higher resolution simulations.

\section*{Acknowledgements}
We would like to thank Marcel Haas for a careful reading of the
manuscript and the anonymous referee for comments that improved its
clarity.  The simulations presented here were run on the Cosmology
Machine at the Institute for Computational Cosmology in Durham as part
of the Virgo Consortium research programme, on Stella, the LOFAR
BlueGene/L system in Groningen, and on Huygens, the Dutch national
supercomputer. This work was supported by an NWO Vidi grant.

\bibliographystyle{mn2e}

\label{lastpage}

\end{document}